\documentclass{emulateapj}
\usepackage{graphicx,amsmath,amssymb}
\usepackage{natbib}
\usepackage{apjfonts}
\bibpunct[, ]{(}{)}{;}{a}{}{,}

\newcommand{\sect}{\S\,}

\begin{document}

   \title{The soft X-ray blast in the apparently sub-luminous GRB\,031203}

   \author{D.~Watson,\altaffilmark{1}
           S.~A.~Vaughan,\altaffilmark{2}
           R.~Willingale,\altaffilmark{2}
           J.~Hjorth,\altaffilmark{1}
           S.~Foley,\altaffilmark{3}
           J.~P.~U.~Fynbo,\altaffilmark{1}
           P.~Jakobsson,\altaffilmark{1}
           A.~Levan,\altaffilmark{2}
           P.~T.~O'Brien,\altaffilmark{2}
           J.~P.~Osborne,\altaffilmark{2}
           K.~Pedersen,\altaffilmark{1}
           J.~N.~Reeves,\altaffilmark{4,5}
           J.~A.~Tedds,\altaffilmark{2}
       and M. G. Watson\altaffilmark{2}
           }
   \altaffiltext{1}{Dark Cosmology Centre, Niels Bohr Institute, University of Copenhagen, Juliane Maries Vej 30, DK-2100 Copenhagen \O, Denmark; darach, jens, pallja, kp @astro.ku.dk}
   \altaffiltext{2}{X-Ray Astronomy Group, Department of Physics and Astronomy, Leicester University, Leicester LE1 7RH, UK; sav2, rw, anl, pto, julo, jat, mgw @star.le.ac.uk}
   \altaffiltext{3}{Dept.\ of Experimental Physics, University College Dublin, Belfield, Dublin 4, Ireland: sfoley@bermuda.ucd.ie}
   \altaffiltext{4}{Laboratory for High Energy Astrophysics, Code 662, NASA Goddard Space Flight Center, Greenbelt, MD 20771, USA; jnr@milkyway.gsfc.nasa.gov}
   \altaffiltext{5}{Universities Space Research Association}

   \begin{abstract}
    GRB\,031203 was a very low apparent luminosity $\gamma$-ray burst (GRB). 
    Coincidentally, it was also the first GRB with a dust-scattered X-ray
    halo. The observation of the halo allowed us to infer the presence of a
    large soft X-ray fluence in the total burst output. It has, however,
    also been claimed that GRB\,031203 was intrinsically sub-energetic,
    representative of a class of spectrally hard, low-energy bursts quite
    different from other GRBs. A careful reanalysis of the available data,
    confirms our original finding that GRB\,031203 had a very large soft
    X-ray component, the time of which can be constrained to within a few
    minutes after the burst, strongly suggesting that while GRB\,031203 did
    indeed have a very low apparent luminosity, it was also very soft. 
    Notions propagated in the literature regarding the uncertainties in the
    determination of the soft X-ray fluence from the halo data and on the
    available constraints from the hard X-ray data are addressed: the
    properties of the scattering dust along the line of sight (grain sizes,
    precise location and the geometry) are determined directly from the high
    quality X-ray data so that there is little uncertainty about the
    scatterer; constraints on the X-ray lightcurve from the \emph{Integral}
    spacecraft at the time of the soft X-ray blast are not complete because
    of a slew in the spacecraft pointing shortly after the burst. Claims
    that GRB\,031203 was intrinsically under-energetic and that it
    represents a deviation from the luminosity--peak energy relation do not
    appear to be substantiated by the data, regardless of whether the soft
    X-ray component is (arbitrarily) declared part of the prompt emission or
    the afterglow. We conclude that the difference between the soft and hard
    X-ray spectra from \emph{XMM-Newton} and \emph{Integral} indicate that a
    second soft pulse probably occurred in this burst as has been observed
    in other GRBs, notably GRB\,050502B.
   \end{abstract}

   \maketitle

\section{Introduction\label{introduction}}

While $\gamma$-ray bursts (GRBs) are no longer as enigmatic as they were
even a few years ago, the ability to use GRBs as a serious tool in cosmology
and an understanding of their basic mechanisms still elude us. Relations
based on the energy release have the potential to resolve these
difficulties.

In particular, the `Amati relation' \citep{2002A&A...390...81A} between the
equivalent isotropic $\gamma$-ray total energy ($E_{\rm iso}$) and the
spectral peak energy $E_{\rm peak}$ in GRBs, has been the focus of
considerable recent work
\citep*[e.g.][]{2005ApJ...627..319B,2005MNRAS.360L..73N,2004ApJ...616..331G}. Only a
single burst apart from GRB\,031203, has extended this relation to very low
luminosities and peak energies \citep[i.e.\ the low luminosity
XRF\,020903,][]{2003astro.ph..9455S}.

It has also been suggested that the total energy in $\gamma$-rays from a GRB
is nearly constant at $\sim10^{51}$\,erg
\citep{2001ApJ...562L..55F}, by correcting for the
opening angle of the putative GRB jet. The determination of the opening
angle is dependent on the time of the break in the lightcurve. This measure
has proved difficult to use or understand because of 1) the difficulty in
deciding the jet break time in lightcurves that are often sparsely-sampled,
contaminated by supernova features, and subject to fluctuations caused by
density variations, and 2) the (few) cases where the total apparent energy
release (equivalent isotropic) is well below this value.

By any measure the apparent isotropic energy output in GRB\,031203 was
extremely low \citep[hereafter W04]{2004ApJ...605L.101W}, and for any
opening angle of the jet, was significantly below the standard energy of
$\sim10^{51}$\,erg for GRBs inferred from jet opening angles (W04).
\citet*[hereafter SLS04]{2004Natur.430..646S} find an isotropic equivalent
energy release of $4\pm1\times10^{49}$\,erg from the \emph{Integral}
20--200\,keV spectrum (an observed fluence of
$2.0\pm0.4\times10^{-6}$\,erg\,cm$^{-2}$). Other bursts (e.g.\ XRF\,020903)
also have apparent energies below $\sim10^{51}$\,erg.

It has been argued by SLS04 and by \citet[hereafter
S04]{2004Natur.430..648S} that GRB\,031203 was representative of a new class
of intrinsically sub-energetic bursts, possessing many of the
characteristics of classical GRBs, but being a thousand times less powerful. 
This claim has far-reaching implications for GRBs. Ambitions to use GRBs as
the most powerful distance indicators in cosmology currently seem to lie
mostly with the $E_{\rm peak}$--$E_\gamma$ relation \citep[similar to the
Amati relation, but using the total collimation-corrected $\gamma$-ray
energy release, $E_\gamma$,][]{2004ApJ...616..331G}, but whatever relation
is used, a low-redshift calibration sample will be essential. If there is a
distinct population of under-energetic bursts, it will clearly need to be
well-described and calibrated differently, especially if this type of burst
dominates the low redshift sample.
 
To suggest that GRB\,031203 was intrinsically sub-energetic and a member of
a new class of such bursts we must answer the question: was the total burst
energy of GRB\,031203 lower than expected compared to other GRBs? Such an
apparently faint burst is expected to be soft according to the Amati
relation. Under the assumption that the emission detected by \emph{Integral}
comprised the entire burst, GRB\,031203 was indeed much fainter than
expected from this relation, since the \emph{Integral} spectrum is hard. The
high value of $E_{\rm peak}$ adopted ($>190$\,keV), was based on the hard
X-ray spectrum of the single pulse detected by the \emph{Integral}
satellite. But as we showed \citep[W04 and][hereafter
V04]{2004ApJ...603L...5V}, the transient dust-scattered X-ray halo
associated with the burst indicates that it was also very rich in soft
X-rays, otherwise the halo observed by \emph{XMM-Newton} could not have been
so bright.

The argument that GRB\,031203 was a member (with GRB\,980425) of a new,
intrinsically under-energetic class of GRBs (SLS04; S04) hinges on the
hardness of the burst.
The fluence in the soft X-ray blast is critical to this discussion.

The \emph{XMM-Newton} data are therefore carefully reanalysed in this
\emph{Letter}. The dominant uncertainties in deriving the fluence are
outlined in \sect\ref{method}. The full spectrum of GRB\,031203 and the
consequences of analysing the complete dataset are presented and discussed
in \sect\ref{discussion}.

A cosmology with $\Omega_{\rm m} = 0.3$, $\Omega_\Lambda = 0.7$ and
H$_0=75$\,km\,s$^{-1}$\,Mpc$^{-1}$ is assumed throughout. Error ranges
quoted are 90\% confidence intervals, unless stated otherwise.

\section{Method and uncertainties \label{method}}

Details of the \emph{XMM-Newton} observations and the initial data analyses
for GRB\,031203 are outlined in V04 and W04. The luminosity of the soft
X-ray blast, inferred from the dust-scattered halo observed by
\emph{XMM-Newton}, is key to the nature of GRB\,031203.  Here, we outline
the procedure used to derive the fluence and analyse the major sources of
uncertainty in this calculation.

A complete model of the X-ray halo was used to find the best-fit parameters,
including the rate of expansion, the width, the total fluence and the flux
decay of the rings. The model produces a two dimensional distribution for
the halo brightness with time and scattering angle for a given energy band.

The fluence of the X-ray blast was inferred from the observed halo fluence
divided by the scattering fraction. The differential scattering fraction as
a function of scattering angle at a given energy is found by integrating
scattering cross-sections over the dust grain size ($a$) distribution up to
the maximum grain size ($a_{\rm max}$) and multiplying by the column density
of dust. The uncertainty in the inferred blast fluence largely reflects the
uncertainties in the scattering dust which is dominated by two things: 1)
the size of the scattering dust column, and 2) the dust grain size
distribution. 

\subsection{The scattering dust column}
It was argued by \citet{2004ApJ...611..200P} and later by SLS04 that the
fluence derived from the X-ray halo could have been overestimated by a
factor of 4.4 in our previous work (V04). This was based on two incorrect
assumptions.

The first was that the individual rings observed in the halo were scattered
by the total dust column along the line of sight (A$_V = 3.6$)\footnote{A
higher A$_V$ means a larger fraction of the X-rays are scattered which in
turn would imply a smaller `blast' fluence for a given observed fluence in
the halo.}, whereas in fact the dust contributing to the rings is confined to
relatively thin sheets of dust at well-defined distances\footnote{The
distance to the scatterer is known from fitting the halo's angular expansion
with time; from geometrical arguments, $D = 2c\tau/\theta^2$, where D is the
distance to the scatterer, $\tau$ is the delay between arrival times of
directly observed and scattered photons, and $\theta$ is the observed angle,
see V04 for more details.} $1395^{+15}_{-30}$\,pc and $868^{+17}_{-16}$\,pc,
see below). Dust that is not contained in these sheets cannot contribute to
the scattered rings and since we use only the X-ray fluence in the rings
themselves to derive the total fluence, other dust along the line of sight
is irrelevant to the calculation of the burst fluence. Even using A$_V=3.6$
as the extreme upper limit to the dust column contained in the sheets does
not change our results by more than a factor of 1.8. At the same time it was
also argued by \citet{2004ApJ...611..200P} that A$_V\sim1$ of the total dust
column actually belongs to the GRB host galaxy, which would leave only
A$_V\sim2.6$ as the upper limit to the dust column available for the
Galactic dust sheets. We find it unlikely that the entire dust column in
this direction is contained in these two sheets. Based on the Galactic
radial dust profiles \citep*{1980A&AS...42..251N}, the most likely value in
the sheets is in fact A$_V\sim2.0$ (V04).

\begin{figure}
 \includegraphics[angle=-90,width=\columnwidth,clip=]{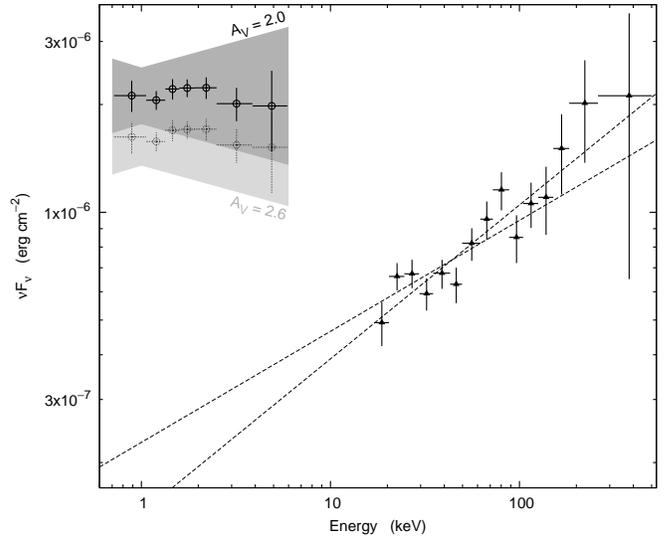}
 \caption{Spectral energy distribution of the pulses detected using the
          dust-scattered X-ray halo and directly with \emph{Integral}.  The
          data from the X-ray halo are plotted as open circles with the
          uncertainty in the correction for the dust scattering plotted as a
          grey butterfly.  The closed triangles represent data from the
          direct observation by \emph{Integral}'s IBIS instrument (SLS04),
          with the 90\% limits to the best-fitting power-law (photon index,
          $\Gamma=1.63\pm 0.06$) plotted as dashed lines---the fluence at
          1\,keV derived from the halo cannot be made consistent with it.
          }
 \label{fig:sed}
\end{figure}

The second misapprehension was that the dust scattering fraction scales
exponentially with A$_V$, whereas the dependence scales with the optical
depth and is therefore only linearly related to the column density
\citep{1986ApJ...302..371M}. The factor 4.4 is the difference in
\emph{optical extinction} between A$_V= 3.6$ and A$_V= 2.0$, not the column
density. This extinction relation is not correct for the X-ray scattering
where the relationship is essentially linear. Since the maximum A$_V$ has
been argued to be $\sim2.6$, the fluence in the burst could only have been
overestimated by at most $2.6/2.0$ which is $\sim30$\%, though as noted
above, this is unlikely.  The effect of using A$_V = 2.6$ to derive the
X-ray burst fluence is illustrated by dotted open circles and a lighter grey
butterfly in Fig.~\ref{fig:sed}.

\subsection{The grain size distribution}
\begin{figure}
 \includegraphics[angle=-90,width=1.0\columnwidth,clip]{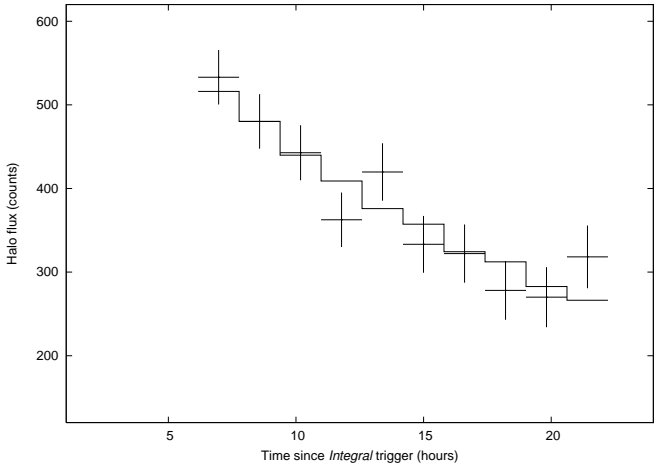}
 \caption{Flux decay of the soft X-ray halo with time with best-fit model
          based on the dust-grain angular scattering profile. Over the
          period of the observation, the total flux in the scattered halo
          decreases as the scattering from increasingly large angles is
          observed.  Increasing the grain sizes makes this decay rate faster
          because larger grains scatter more efficiently at smaller angles. 
          The maximum of the grain size distribution ($a_{\rm max}$),
          dominates the scattering profile.  We can therefore fit to the
          measured decay of the halo to find $a_{\rm max}$. In this case,
          $a_{\rm max} = 0.50\pm0.03\,\mu$m.
         }
 \label{fig:halo_lightcurve}
\end{figure}

Because we possess time-resolved data for the evolution of the X-ray halo,
the shape of the angular scattering response function for the dust (i.e.\
the way the scattered flux falls off with the scattering angle,
Fig.~\ref{fig:halo_lightcurve}) is strongly constrained. The largest grains
always dictate this angular scattering response function, allowing us to fit
the differential scattering cross-sections to the observed flux in the halo
as a function of scattering angle, with $a_{\rm max}$ as a free parameter.
This allows us to say that $a_{\rm max}=0.50\pm0.03\,\mu$m along this line of sight.

In our original analysis (V04) a single grain size was used. Here we assume
a distribution proportional to $a^{-3.5}$, which gives a good representation
of the optical extinction curve and is similar to that observed in X-ray
scattering for Galactic sources
\citep*{1977ApJ...217..425M,1986ApJ...302..371M,1995A&A...293..889P}

Running the model with different values of the power law index of the grain
size distribution, it is clear that values below $-4.0$ yield very large
total scattering fractions per A$_V$, ($>12$\%), well above any observed
value \citep{2003ApJ...598.1026D}. Even using the steep value of $-4.0$
implies a fluence only $\sim33$\% smaller than the our best estimate.

The results from this analysis are consistent with our previous estimate
(V04) that used a single grain size, based on the dust scattering
efficiencies found for Galactic X-ray halo sources \citep{1995A&A...293..889P}.  In other
words, a similar Galactic source halo would have close to the same
brightness for its central source as we infer for GRB\,031203.

The scattering efficiency is not very sensitive to variations in the details
of the physical grain model.

\subsubsection{Dust scattering efficiency}
The dust model of \citet{2001ApJ...548..296W} which has been used
\citep[e.g.][]{2004ApJ...617..987D} to convert optical extinction
(A$_V$) to X-ray scattering factor ($\tau_{\rm sca}$), gives a scattering
factor that is consistently 2--4 times larger, over the 0.7--3\,keV range,
than empirically determined from Galactic sources using \emph{ASCA}, ROSAT
and \emph{Chandra}
\citep{1994ApJ...436L...5W,1995A&A...293..889P,2002ApJ...581..562S,2003ApJ...594L..43Y}.
Using this model, the fluence in the X-ray blast would be lower by a factor
of $\sim3$. However, as this model does not agree with the observational
comparison of $\tau_{\rm sca}$ and A$_V$ \citep[see Fig.~11
of][]{2003ApJ...598.1026D}, we have continued to use the empirically
determined value from \citet{1995A&A...293..889P}. It is worth noting that
the $\tau_{\rm sca}$--A$_V$ relation is strongly dependent on the grain size
distribution (see above), so that these are not independent sources of
uncertainty.

\section{Results}
Our reanalysis results in a 1\,keV fluence density of
$1320\pm260$\,ph\,cm$^{-2}$\,keV$^{-1}$
($2.1\pm0.4\times10^{-6}$\,erg\,cm$^{-2}$\,keV$^{-1}$), a factor of ten
above an extrapolation of the \emph{Integral} power-law spectrum to 1\,keV
(which has a 1\,keV fluence density of $110\pm20$\,ph\,cm$^{-2}$\,keV$^{-1}$
and a photon index, $\Gamma=1.63\pm0.06$, Fig.~\ref{fig:sed}).  The
uncertainties quoted for the X-ray halo data include the measurement error
and the uncertainties related to the halo modelling. Given the size of these
uncertainties and the fact that they are based on direct observation along
this line of sight, we are forced to conclude that it is unlikely that the
1\,keV fluence of the blast could have been substantially different.

The analysis of the halo expansion was also improved by allowing the time of
the X-ray blast to be a free fit parameter, Gaussian profiles were fit to
the halo at different times to improve the radial size estimates, and the
model fit was integrated over each time bin. We find results consistent with
our previous work.  The time of the blast was $600\pm700$\,s after the
beginning of the burst detected by \emph{Integral}. The distances to the
scatterers of $1395^{+15}_{-30}$\,pc and $868^{+17}_{-16}$\,pc are some of
the most accurately known distances to any object beyond about 50\,pc, with
a total uncertainty of only $\lesssim2\%$ at $\sim1$\,kpc.

\section{Discussion\label{discussion}}

The peculiar SED of the complete dataset, points to the fact that
\emph{Integral} and \emph{XMM-Newton} observed different events in
GRB\,031203. A natural interpretation of these data is that there were two
pulses in GRB\,031203: one detected by \emph{Integral}, with a hard spectrum
peaking at or above $\sim190$\,keV, and a second pulse with a much softer
spectrum, detected by \emph{XMM-Newton} via its dust-scattered halo.

It is expected that \emph{Integral}'s IBIS instrument \emph{should}, in its
lowest energy channels, have detected the harder X-rays associated with such
a powerful soft X-ray blast (Sazonov, priv. comm.). However, the lightcurve
limits obtained by \emph{Integral} are incomplete. Long ($\lesssim40$\,s)
data gaps exist due to a $\sim100$\,s slew of the satellite. The slew
occurred less than 300\,s after the initial pulse.  The IBIS data cannot
therefore be used to place useful limits on the soft flux in the burst. (It
may however be used to place limits on the timing of the X-ray blast).

Many bursts exhibit multiple pulses often accompanied by a strong softening
of the spectrum, e.g.\ GRB\,960720 or GRB\,970228
\citep{2000ApJS..127...59F} or GRB\,020410 \citep{2004A&A...427..445N}. The
most striking case so far appeared during the preparation of this paper; the
detection of a massive soft X-ray flare in GRB\,050502B
\citep{2005astro.ph..6130B} starting $\sim500$\,s after the initial
$\gamma$-ray pulse, and lasting $\sim500$\,s. The fluence in the soft X-ray
flare was comparable to the first $\gamma$-ray pulse, which had a hard
spectrum. (Indeed, the photon spectral index, $\Gamma=1.6$, was very similar
to that observed in the $\gamma$-ray pulse of GRB\,031203.) The consistency
between the features observed in GRB\,050502B and those inferred here from
the X-ray halo of GRB\,031203, reinforces the interpretation that there were
two very different pulses in GRB\,031203.

\subsection{Afterglow or prompt emission?}
The complete data show that although GRB\,031203 was very faint (W04), there
is no reason to suppose that it was anomalously so---it is more luminous
than XRF\,020903, for instance and probably of comparable luminosity with
XRF\,030723 \citep{2004ApJ...609..962F,2004ApJ...612L.105T}. The key issue
is therefore whether it deviates significantly from the `Amati relation',
i.e.\ whether it was spectrally hard. The interpretation of the
event---prompt emission, highly unusual afterglow, or reverse shock---while
interesting speculation (SLS04), is not relevant to whether or not the burst
was unexpectedly faint.  The comparison is an observational one, i.e.\
\citet{2002A&A...390...81A} used the emission detected by the
\emph{BeppoSAX} burst monitor and Wide Field Camera (WFC).  To compare with
these bursts in a meaningful way, we must use the same observational
criteria and must include the soft X-ray blast in the calculation of the
total luminosity, since its fluence or minimum flux would have been detected
by the WFC \citep{2002A&A...390...81A}. The consideration of whether certain
parts of the emission should or could be considered as afterglow or prompt
emission is irrelevant for this comparison, which is an observational one,
based on the criteria for the sample selection. Based purely on the
\emph{Integral} data, GRB\,031203 appears to be one of only two significant
outliers from this relation (the other being GRB\,980425). However, when we
include the \emph{XMM-Newton} data, the X-ray (2--30\,keV) to $\gamma$-ray
(30--400\,keV) fluence ratio is $S_{\rm X}/S_\gamma = 1.8^{+0.4}_{-0.9}$,
which indicates that GRB\,031203 was probably an X-ray flash, and certainly
at least X-ray rich. This implies not only that the lower bound to the total
X- and $\gamma$-ray fluence in the burst was roughly twice the
$2\times10^{-6}$ given by SLS04, but more importantly, that the peak energy
of the total burst (if this is a well-defined concept in this case) was
likely very low as originally concluded in W04. Taking this into account, we
conclude that there is no compelling evidence in the GRB energetics to
suggest GRB\,031203 was intrinsically under-energetic.

In support of the argument that GRB\,031203 was a cosmic analogue of
GRB\,980425, it was suggested that the shape of the prompt emission in was
the same in both bursts (single pulse and FRED shape, S04). Since
GRB\,031203 could certainly have possessed multiple pulses, this suggestion
is not compelling.

The luminosity of the X-ray afterglow at about one day is also very faint
($9\times10^{42}$\,erg\,s$^{-1}$ at 10\,hr, W04). The X-ray afterglow is,
however, still two orders of magnitude \citep{2005ApJ...625L..91R} brighter
than predicted in the sub-energetic model proposed by S04 and both this and
the low energy inferred from radio calorimetry ($1.7\times10^{49}$\,erg,
S04) can be readily explained in a standard energy, off-axis model
\citep{2005ApJ...625L..91R}, which suggests an intrinsic peak energy for the
burst of a few hundred keV, an order of magnitude above the total
observed XMM-\emph{Newton}+\emph{Integral} value.

\acknowledgments
We acknowledge benefits from collaboration within the EU FP5 Research
Training Network, `Gamma-Ray Bursts: An Enigma and a Tool'. This work was
also supported by the Danish Natural Science Research Council (SNF).

\end{document}